\documentclass[a4paper,twocolumn,11pt, accepted=2022-07-23]{quantumarticle}

\pdfoutput=1

\usepackage{amsfonts,amsmath,amssymb,amsthm,graphicx, bm,upgreek}
\usepackage{physics}
\usepackage{color}
\usepackage{ifthen}
\usepackage[toc,page]{appendix}
\usepackage{dsfont}
\graphicspath{{figs\_QSM1/}}

\newcommand{\code}{\texttt}

\def\pindex{p}
\def\qindex{q}
\def\Tparam{\hbar}
\def\Lparam{L}
\def\Kparam{K}
\def\kparam{k}
\def\Gparam{G}

\include{macros}

\begin{document}

\title{Observability of fidelity decay at the Lyapunov rate in few-qubit quantum simulations}

\author{Max D. Porter}
\email{porter42@llnl.gov}
\affiliation{Fusion Energy Sciences Program, Lawrence Livermore National Laboratory}
\author{Ilon Joseph}
\affiliation{Fusion Energy Sciences Program, Lawrence Livermore National Laboratory}


\maketitle

\begin{abstract}
	In certain regimes, the fidelity of quantum states will decay at a rate set by the classical Lyapunov exponent. This serves both as one of the most important examples of the quantum-classical correspondence principle and as an accurate test for the presence of chaos.  While detecting this phenomenon is one of the first useful calculations that noisy quantum computers without error correction can perform [G. Benenti et al., Phys. Rev. E 65, 066205 (2001)], a thorough study of the quantum sawtooth map reveals that observing the Lyapunov regime is just beyond the reach of present-day devices. We prove that there are three bounds on the ability of any device to observe the Lyapunov regime and give the first quantitatively accurate description of these bounds: (1) the Fermi golden rule decay rate must be larger than the Lyapunov rate, (2) the quantum dynamics must be diffusive rather than localized, and (3) the initial decay rate must be slow enough for Lyapunov decay to be observable. This last bound, which has not been recognized previously, places a limit on the maximum amount of noise that can be tolerated.  The theory implies that an absolute minimum of 6 qubits is required. Recent experiments on IBM-Q and IonQ imply that some combination of a noise reduction by up to 100$\times$ per gate and large increases in connectivity and gate parallelization are also necessary. Finally, scaling arguments are given that quantify the ability of future devices to observe the Lyapunov regime based on trade-offs between hardware architecture and performance.
\end{abstract}

\section{Introduction}

\subsection{Motivation}

Understanding chaotic dynamical systems is essential for understanding the world around us.
Because chaotic systems are typically modeled through numerical computation, they are one of the primary application areas for scientific computing.
Examples from classical physics include simulations of molecular dynamics \cite{magann2021digital}, fluid dynamics \cite{gaitan2020finding, gaitan2021finding}, plasma physics \cite{dodin2020applications, shi2021simulating}, Monte Carlo methods, and gravitational N-body problems. 
Examples from quantum physics include non-equilibrium condensed matter physics \cite{le2016many}, chemistry \cite{mcardle2020quantum}, nuclear physics \cite{de2021quantum, holland2020optimal}, and lattice gauge theory \cite{martinez2016real}. 
Since quantum computing offers a potential acceleration of many important calculations within scientific computing \cite{montanaro2016quantum, childs2010quantum, montanaro2015quantum, tilly2021variational}, chaotic dynamical systems stand to be one of the most important quantum computing application areas.
In fact, because chaotic systems are provably difficult to simulate, recent claims of achieving quantum supremacy \cite{boixo2018characterizing, arute2019quantum} crucially rely on the exponential difficulty of simulating chaotic quantum circuits on a classical computer. Simulating quantum chaotic dynamics has even been proposed as the most qubit-efficient application for reaching useful quantum advantage \cite{babbush2021youtube}.

Simulating \textit{classical} chaotic systems on a quantum computer comes with a particular challenge: how can a nonlinear dynamical system be simulated in a quantum computer that can only efficiently perform linear operations?
Quantum computers were originally proposed \cite{feynman1982simulating, manin1980computable} because they are well-adapted to simulating quantum mechanical systems. To realize this, a number of quantum algorithms for quantum Hamiltonian simulation have been proposed that can provide an exponential speedup over direct simulation \cite{lloyd1996universal}.
Therefore a natural approach for classical Hamiltonian systems is to simulate their quantized versions and find ways to extract classically-relevant information.
In fact, it was recognized early on that this technique can accelerate the computation of useful dynamical quantities, such as the Lyapunov exponent \cite{benenti2001efficient, benenti2004quantum}. This is possible even without using many qubits to approach the limit of classical dynamics, raising the possibility of observing one of these quantities in few-qubit quantum simulations.

Simulating non-Hamiltonian classical systems requires other approaches, not employed in this paper.
Techniques for exactly simulating such systems have recently been proposed \cite{joseph2020koopman, dodin2020applications, liu2020efficient, lloyd2020quantum, engel2021linear, dodin2021quantum}. 
The key is to embed the nonlinear system within an infinite-dimensional linear system, and then determine a finite-dimensional approximation that has sufficiently high accuracy for the purposes at hand.
Then, if the linear system is unitary, as in the Koopman-von Neumann approach \cite{joseph2020koopman, dodin2020applications}, one can directly use Hamiltonian simulation algorithms.
If it is not unitary, as in the Carleman linearization approach \cite{liu2020efficient, engel2021linear}, one can still use the quantum linear solver algorithm \cite{harrow2009quantum, childs2017quantum} to propagate the state forward in time.

The speedup predicted for quantum simulation of any system applies only to information that can be efficiently extracted. For dynamics this means collective properties like the Lyapunov exponent, localization length, and diffusion coefficient. The Lyapunov exponent is extracted from the fidelity of a quantum algorithm in the presence of noise, making it relevant in the NISQ era. If the algorithm is exponentially efficient compared to classical methods, for instance by use of the quantum Fourier transform, the Lyapunov exponent inherits an exponential speedup \cite{benenti2004quantum}.

A near term objective then for the quantum simulation of classically chaotic dynamics is to observe the Lyapunov exponent on a noisy few-qubit quantum device. This doubles as a signature of quantum chaos, and one that is more accessible and efficient to detect than more traditional measures such as level spacing statistics.
This objective faces three limitations: (1) the Fermi golden rule (FGR) decay rate, which scales with the square of the noise, must be larger than the Lyapunov exponent, (2) the Lyapunov exponent must be large enough to avoid a phase transition to localization, and (3) the noise must be small enough that the initial decay rate allows enough time steps of the fidelity decay to be obtained. Together these conditions form a triangular region in phase space such that only experiments with the right noise magnitude, Lyapunov exponent, and number of qubits will fall inside this triangle and be able to observe the Lyapunov exponent. The minimum number of qubits for this region to have non-zero area will be shown to be six.

When targeting this objective using near-term quantum computing hardware platforms, one must consider the trade-off between architecture and gate error rates for reaching the threshold decay rate of the first time step. Recent experiments on the open access IBM-Q superconducting platform and reported data from IonQ, two commercial platforms known for their high qubit fidelities, show that neither can presently observe Lyapunov decay. This could change if either reduces the error per two-qubit gate by a factor of about ten. Alternatively, both platforms could pursue architectural improvements, with IBM-Q needing to improve on the low qubit connectivity seen in almost all superconducting quantum devices and IonQ needing to perform gates in parallel in less time than the same gates would take in serial. Both obstacles have seen progress, and could lower the needed error reduction in these platforms to a factor of one to three, potentially entering the Lyapunov regime. Scaling arguments show that extending to more than six qubits is relatively easy, with the increased size of the Hilbert space lowering the $1/N$ fidelity limit, which increases the number of useful time steps and therefore the allowable noise. With the ideal architecture more qubits may even allow larger error per gate than for fewer qubits, making architectural improvements particularly important for scaling up to reach quantum advantage.

\subsection{Quantum Maps}
Classical maps are model systems of deterministic, often Hamiltonian, chaos and nonlinear dynamics, and their quantized counterparts are called quantum maps. Just as classical maps use discrete time steps to achieve rich dynamics at low computational cost, quantum maps are Floquet systems with trivial time ordering, leading again to discrete time steps and rich yet efficient dynamics at low gate depth. In the future, scalable, error-corrected devices will simulate quantum maps in their classical limit, resolving classical phase space structures larger than $\hbar/J_0\sim1/N$ for classical action $J_0$ and Hilbert space dimension $N$ \cite{benenti2001efficient}. In the present and near term however quantum maps are an excellent tool for exploring the simulation of both quantum and classical chaos, while using minimal resources. Quantum maps have a numerically precise yet dense evolution operator and a simple classical correspondence principle. Coupled quantum maps are akin to spin chains and allow the study of many-body physics without need for trotterization \cite{notarnicola2020slow}.

Specific algorithms using standard gate sets have been proposed for digital quantum simulations of quantized versions of the standard map \cite{georgeot2001exponential}, sawtooth map \cite{benenti2001efficient}, kicked Harper map \cite{levi2004quantum}, tent map \cite{frahm2004quantum}, and baker's map \cite{schack1998using}. 

The quantum baker’s map could be potentially interesting to study because it is the most efficient to implement, requiring only quantum Fourier transforms.  However, one may need to generalize the form of the baker’s map in order to have enough free parameters to avoid localization and ensure the observability of the Lyapunov regime.

This paper instead continues the pioneering work of Benenti et al. \cite{benenti2001efficient, benenti2002quantum, benenti2003dynamical, benenti2004quantum} in using the second-most resource efficient quantum map, the quantum sawtooth map, which has been shown to have rich dynamics. It consists of four steps with $O(n^2)$ gates each, before accounting for qubit topology or converting to gate depth. Theoretical studies looking at the quantum sawtooth map have often used $\sim$10 qubits \cite{benenti2004quantum, wang2004stability}, while an experimental realization of this system on the IBM-Q platform has demonstrated its present feasibility on three qubits \cite{pizzamiglio2021dynamical}. To bridge theory and experiment, this work seeks to find the smallest system size and largest noise for which the Lyapunov exponent is still observable in this system. 

The most thorough previous study of quantum simulation of the quantum sawtooth map \cite{benenti2004quantum} missed several key observations concerning the Lyapunov exponent which we clarify and combine here. First, they did not mention the correct minimum noise condition for observing the Lyapunov exponent, which had been discovered several years before \cite{jacquod2001golden} and will form bound (1) in Sec.~\ref{sec:three_bounds} of this paper. Second, their estimate of the border to localization was imprecise and has been improved to form bound (2). Third and lastly, they did not recognize the early time regime which is crucial for quantifying the maximum allowable noise in bound (3). These details are necessary in the context of few-qubit quantum simulation.

\subsection{ Loschmidt echo}
Quantum maps have a long history as a tool for studying quantum chaos, quantum computing, and especially the Loschmidt echo (LE or ``fidelity"). The nuanced relationship between these three has been outlined from various points of view in several review papers \cite{jacquod2009decoherence, gorin2006dynamics, goussev2012loschmidt}. This fidelity typically measures the overlap of a state after time evolution under two different quantum Hamiltonians, one with and one without a perturbation. In order to match experiments, this work will use a ``two-way" fidelity that is the overlap of two such evolutions that are independently perturbed. This overlap decays at a gradual and context-dependent rate in time, so it is useful also for linearized classical flows \cite{eckhardt2002echoes}.

Asher Peres \cite{peres1984stability} initiated the study of how dynamics impacts fidelity decay in quantum systems and showed how this understanding can shed light on the classical-quantum correspondence principle for chaotic systems. This inspired a flurry of work decades later laying out regimes of quantum fidelity decay, most importantly for chaotic systems a Lyapunov exponent decay rate at large perturbations or weak chaos \cite{jacquod2009decoherence, jalabert2001environment, jacquod2001golden}. This can be concealed by early time and late time decay regimes \cite{jacquod2009decoherence} and in certain cases by an oscillating decay rate at the transition to Lyapunov-rate decay \cite{wang2004stability, ares2009loschmidt, garcia2011loschmidt}. Random matrix theory, semiclassical path integrals, and other methods have been employed to explain these regimes as well as more subtle effects \cite{gorin2006dynamics, jacquod2009decoherence}. As applied to quantum computing, an important distinction is between static and random errors in the Hamiltonian, analogous at the gate level to coherent and incoherent errors in real quantum hardware \cite{sutherlandprivate}. Whereas random errors give exponential decay with linear dependence on the number of gates in the exponent, static errors can cause quadratic dependence for large numbers of gates \cite{frahm2004quantum}. In this work simulations only consider random error for simplicity, but static imperfections should be studied in the future.

Fidelity decay rates may serve as an especially useful signature of quantum chaos in quantum simulation. Where energy level statistics need many independent level spacings to resolve the distribution, fidelity decay relies on the more sensitive eigenstates. Eigenstate deformation was recently shown to be a more sensitive measure of quantum chaos than level statistics in a many-body system, and to be sensitive at increasingly small system sizes as the strength of perturbation grows \cite{pandey2020adiabatic}. This is somewhat supported by experiment: the level statistics approach was used to detect a quantum-chaotic to many-body-localized (MBL) transition in a nine-qubit specialized emulator \cite{roushan2017spectroscopic}, while recent work used the fidelity decay rate to detect a diffusive to localized transition in a three-qubit digital quantum simulator \cite{porter2021slowed}. The fidelity is also a more accessible if coarser tool than measuring the energy spectrum and could simplify experiments.

\subsection{Overview of Contents }
In the next section, the sawtooth map is introduced in both classical (Sec.~\ref{sec:CSM}) and quantum (Sec.~\ref{sec:QSM}) form and its key properties are described. An expression is derived in Sec.~\ref{sec:localization} for the dynamical localization length in this system as well as a threshold for its observation. In Sec.~\ref{sec:effects_of_noise} the system's noise-induced fidelity decay regimes are described, and related to previous work. 
In Sec.~\ref{sec:three_bounds} formulas are derived for the three bounds restricting Lyapunov fidelity decay. Then in Sec.~\ref{sec:requirements} these parameter bounds are combined with numerical results, recent experiments, and reported fidelities to establish lower bounds for the necessary system size and reduction in noise relative to IBM-Q and IonQ's present day hardware in order to achieve a first observation of Lyapunov fidelity decay.
Section~\ref{sec:conclusion} summarizes and provides some closing thoughts.

\section{The Sawtooth Map} \label{sec:sawtooth_map}

To begin, the classical and quantum sawtooth maps are defined, which are classically chaotic and quantum chaotic respectively. \cite{lakshminarayan1995quantum} This map is chosen for its efficient simulation of Hamiltonian chaos \cite{porter2021slowed}. Expressions are given for the classical Lyapunov exponent and diffusion coefficient, which inform the analysis of the quantum dynamics.

\subsection{Classical Sawtooth Map (CSM)} \label{sec:CSM}

The classical sawtooth map (CSM) is governed by the periodically driven Hamiltonian
\begin{equation}
	\tilde H_{CSM}=\frac{J^2}{2I} - \sum_n \tilde K \frac{\theta^2}{2} \delta(t - n \tau) \text{ for } \theta \bmod 2\pi
\end{equation}

\noindent where $(J,\theta)$ are conjugate action-angle variables, $\tau$ is the driving period, $I$ is a constant moment of inertia, and $\tilde K$ is the kicking parameter with units of action. The sum $n$ is over all integers. It is convenient to remove units from the equation, so the variables are transformed as $J=J' I/\tau, t=t' \tau, \tilde K=K' I/\tau, H=H' I/\tau^2$ then the primes are dropped to get
\begin{equation}
	H_{CSM}=\frac{J^2}{2} - \sum_n K \frac{\theta^2}{2} \delta(t - n) \text{ for } \theta \bmod 2\pi
	\label{eq:H_CSM}
\end{equation}
with dimensionless kicking parameter $K$. Integrating Hamilton's equations of motion over one period gives the map
\begin{equation}
	\begin{array}{l}
		J_{n+1}=J_n+K \theta_n \,\text{ mod } 2\pi \Lparam \, (-\pi L \leq J < \pi L)\\
		\theta_{n+1}=\theta_n+J_{n+1} \,\text{ mod } 2\pi \, (-\pi \leq \theta < \pi)
	\end{array}
\label{eq:map_CSM}
\end{equation}
where $\Lparam$ is a positive integer to ensure no discontinuous behavior in the second (free evolution) equation. In the limit $\Lparam\rightarrow \infty$ the map's manifold shifts from a torus to a cylinder. The nonlinearity of the map arises from its modulo operation. While its dynamics for non-integer $-4<K<0$ are quasi-integrable (the Lyapunov exponent is zero while anomalous diffusion occurs), they become chaotic for $K>0$ and $K<-4$. The maximal Lyapunov exponent describing the strength of the chaos is
\begin{flalign}
	\lambda(K)=\ln[(2+K+\text{sgn}(K)\sqrt{K^2+4K})/2] \nonumber \\
	\approx \begin{cases}
		\ln[\Kparam+2 + O(K^{-1})] \\
		\qquad\qquad\qquad\qquad\qquad\, \text{ for } \Kparam \gg 1 \\
		\Kparam^{1/2} - \frac{1}{24}\Kparam^{3/2} + \frac{1}{4} \Kparam^2 - O(\Kparam^{5/2}) \\
		\qquad\qquad\qquad\qquad \text{ for } 0 < \Kparam \ll 1\\
	\end{cases}
	\label{eq:lyapunov}
\end{flalign}
\cite{benenti2004quantum}. For this paper we choose to restrict the analysis to $K>0$.

Diffusion in classical chaotic systems can be understood as that of a random walk in the action $J$, where the probability density function obeys the Fokker-Planck equation. The main result is that the CSM diffusion coefficient is
\begin{flalign}
	D_\Kparam \approx \begin{cases}
		(\pi^2/3) K^2 \qquad \text{ for } \Kparam>1 \\
		3.3 \Kparam^{5/2} \quad \text{ for } 0<\Kparam<1 \\
	\end{cases}
	\label{eq:diff_coeff}
\end{flalign}
where the first case comes from a random phase approximation, and in the latter case trajectories stick to broken cantori which slows diffusion \cite{benenti2004quantum}.

\subsection{Quantum Sawtooth Map (QSM)} \label{sec:QSM}

The quantum sawtooth map (QSM) can be straightforwardly derived from the classical Hamiltonian of Eq. \ref{eq:H_CSM}. The action variable $J$ is quantized by enforcing the canonical commutation relation $[\hat \theta, \hat J]=i \Tparam$, where $\Tparam$ is the dimensionless Planck's constant given by $\hbar_{phys} = \Tparam'  I/\tau$ then dropping the prime. For computability $\theta$ is then discretized in a computational basis. This limits $J$ to a finite number of  values, as the size of the two bases must be equal for transformations between them to be possible. The operator eigenvalues are then
\begin{equation} \label{eq:quantize_vars}
	\begin{array}{l}
		\hat J \ket{\pindex} = \Tparam \pindex \ket{\pindex}; \quad \pindex = -N/2,...,(N-1)/2 \\
		\hat \theta \ket{\qindex} = 2 \pi \qindex / N \ket{\qindex}; \quad \qindex = -N/2,...,(N-1)/2 
	\end{array}
\end{equation}
where $N$ is the chosen basis size, which in the context of quantum computing is naturally chosen to be $N=2^n$ for $n$ qubits. States $\ket{\qindex}$ and $\ket{\pindex}$ will be referred to as the canonical position and momentum eigenstates respectively.

The quantized Hamiltonian results in a quantum evolution propagator over each period given by
\begin{flalign}
		U_{QSM} &= \hat{\mathcal{T}} \exp(-i \int_0^1 H_{QSM} dt /\Tparam) \nonumber \\
		&= U_{kin} U_{pot} \\
		U_{pot} &= \exp(i \kparam (\beta \hat q)^2 /2) \nonumber \\
		U_{kin} &= \exp(-i \Tparam {\hat p}^2 /2) \nonumber
	\label{eq:QSM_def}
\end{flalign}
where $\hat{\mathcal{T}}$ is the time-ordering operator, $\kparam \equiv \Kparam/\Tparam$ is the quantum kicking parameter, and $\beta \equiv 2\pi/N$ for N basis states. This single-period propagator is often called a Floquet operator, but since it corresponds to a classical map it is also referred to as a quantum map. Quantum map propagators are particularly simple to calculate as the delta-function in the potential makes time ordering trivial. The eigenstates of the quantum map are also eigenstates of the so-called Floquet Hamiltonian, the matrix logarithm of the time-averaged evolution operator, so one can refer to them as Floquet eigenstates or quasienergy eigenstates. Their phase evolution controls the evolution of the system.

In the quantum map there are two key dimensionless parameters $\kparam = \Kparam/\Tparam$ and $\Tparam$. Matching the periodicities of the classical and quantum systems further requires $\Delta J = 2\pi \Lparam M = \Tparam N$, where $\Lparam$ and $M$ are positive integers and $M=1$ is set without loss of generality. This restricts Planck's constant to $\Tparam = 2\pi \Lparam /N$. The classical limit of the QSM is achievable only in the many-qubit limit $N \rightarrow \infty$, causing $\Tparam \rightarrow 0$ and $\kparam \rightarrow \infty$ while keeping the classical parameter $\Kparam = \kparam \Tparam$ constant. In this study $\Lparam=1$ is chosen to maximize the chance of observing a Lyapunov decay rate, based on the bounds in Sec.~\ref{sec:lyap_bounds}. Then $\Kparam$ (controlling $\lambda(\Kparam)$) and noise amplitude $\sigma$ (see Sec.~\ref{sec:effects_of_noise}) are varied to explore the dynamics. A range of system sizes relevant to present day and near term quantum devices are considered, $3 \le n \le 12$. Note that making $\Lparam$ even would allow $\Lparam/N$ to simplify and reduce the pseudorandomness of the phase operators, disrupting the chaotic dynamics.

\section{Dynamics} \label{sec:dynamics}

\subsection{Dynamical Localization} \label{sec:localization}

In the field of quantum chaos, quantum analogs of classically chaotic systems are well-known to diffuse classically until reaching a steady state exponential localization of the wave function
\begin{equation}
	P_\pindex = |\braket{\pindex}{\psi}|^2 \approx \frac{1}{\ell} \exp(- \frac{2|\pindex-\pindex_0|}{\ell}),
\end{equation}
for initial momentum $\ket{\pindex_0}$, under certain conditions. The localization length $\ell$ is reached after the Heisenberg time $\tau_H \approx \ell$ \cite{benenti2004quantum}. The heuristic explanation is that initially there is classical diffusion of strength $D_{\Kparam}$, but this transitions at the Heisenberg time to coherent oscillations of frequencies $\omega = \Delta E / \hbar$ between (quasi)energy eigenstate pairs with energy differences $\Delta E$. The Heisenberg time or ``break" time \cite{benenti2004quantum} is defined as the inverse mean energy level spacing $\tau_H = 2 \pi \hbar / \Delta E_\text{ave}$ \cite{shepelyansky2020ehrenfest,vsuntajs2020quantum} when these oscillations begin to dominate. The randomness of these pair oscillations causes a net effect of localization around the initial state. The balance of these processes is captured by the approximation
\begin{flalign}
	\ell &\approx D_{\Kparam}/\Tparam^2  \nonumber \\
	&\approx \begin{cases}
		(\pi^2/3) \kparam^2 \qquad\qquad\qquad\quad\, \text{ for } \Kparam>1 \\
		3.3 \kparam^{5/2} (2\pi \Lparam/N)^{1/2} \quad \text{ for } 0<\Kparam<1
	\end{cases}
\end{flalign}
using Eq.~\ref{eq:diff_coeff} and $\Kparam = \kparam* 2 \pi \Lparam/N$ \cite{benenti2004quantum, benenti2003dynamical}. In practice dynamical localization is reflected in localized Floquet eigenstates, with the Floquet evolution only including Floquet eigenstates in proportion to their overlap with the initial condition. For $0<\Kparam<1$ there may also occur a second localization regime with slower scaling in $\kparam$ as seen in other quantum maps \cite{borgonovi1998localization, casati1999quantum, prange1999quasiclassical}. This has not been studied for the quantum sawtooth map, but would be in the strongly localized regime and unlikely to affect the border to diffusion below.

Using the expressions for $\ell$ and for the localized distribution, the condition for observing any localization should be that $P_\pindex$ has decayed by less than $1/2$ at the edge, or else the overlap of the tails would cause full diffusion. This gives the condition
\begin{flalign}
	1/2 &> \exp(-N/\ell)  \nonumber \\
	 \ell &< N/\ln(2) \nonumber \\
	 \kparam &< \kparam_\text{loc}  \nonumber \\
	& \equiv \begin{cases}
		\sqrt{\frac{3}{\ln(2) \pi^2} N} \approx 0.66 N^{1/2} \quad \text{ for } \Kparam>1 \\
		\left(\frac{1}{3.3 \sqrt{2\pi} \ln(2)} \frac{N^{3/2}}{\Lparam^{1/2}} \right)^{2/5} \approx 0.50 N^{3/5} \Lparam^{-1/5} \\
		\qquad \qquad \qquad \qquad\quad \text{ for } 0<\Kparam<1
	\end{cases}
\label{eq:loc_condition}
\end{flalign}
 for localization, alternately written as
\begin{equation}
	\Kparam < \Kparam_\text{loc} \approx
	\begin{cases}
		4.16 \Lparam N^{-1/2} \qquad\qquad \text { for } \Kparam>1 \\
		3.12 \Lparam^{4/5} N^{-2/5} \quad \text{ for } 0<\Kparam<1.
	\end{cases}
	\label{eq:loc_condition_K}
\end{equation}

Note that the QSM becomes a quantum cat map when $\Kparam/\Lparam \in \mathbb{Z}$ ($2 \pi \kparam/N \in \mathbb{Z}$). In such cases the generic behaviors of localization and diffusion are replaced by periodic behavior with regular structures in phase space. Cat maps arise when the potential energies $\arg(U_{pot})=\Kparam/\Lparam * \beta \qindex^2 /2 \text{  mod } 2\pi$ are not pseudorandom over $\qindex$. However even small perturbations can restore the generic QSM behavior, so the large noise considered in this paper can make cat maps' periodic behavior difficult to observe in practice \cite{lakshminarayan1995quantum}.

\subsection{Noise} \label{sec:effects_of_noise}

\subsubsection{Noise Model and Fidelity} \label{sec:noise_models}

For this paper a simple parameter noise model is used, where the quantum kicking strength $\kparam$ is perturbed at each step by $\kparam \rightarrow \kparam + \Delta \kparam$. The noise is random and drawn from a normal distribution with standard deviation $\sigma$, where the PDF is given by
\begin{flalign}
	p(\Delta \kparam) = \frac{1}{\sqrt{2\pi} \sigma} \exp(-\frac{\Delta \kparam^2}{2 \sigma^2}).
	\label{eq:k-noise}
\end{flalign}
This noise model easily extends to the classical map by perturbing the classical kick $\Kparam=\kparam \Tparam$ with standard deviation $\epsilon=\sigma \Tparam$.

Below, the effect of noise on the rate of fidelity decay of the quantum system is studied. The fidelity of a noisy unitary evolution $U_\sigma$ can be measured by
\begin{equation}
	f(t) = |\bra{\psi} U_{\sigma'}^{-t} U_{\sigma}^{t} \ket{\psi}|^2
\end{equation}
which is also known as the Loschmidt echo. In words, an initial state $\ket{\psi}$ is evolved for a discrete number of time steps $t$ by a unitary process $U_\sigma$ formed from an ideal $U$ and a noisy process of magnitude $\sigma$. The inverse operation is then attempted with a different realization of the random noise in order to recover the original state, with the fidelity of success $f(t)$ measured in the basis of the initial state. (For non-unitary, Markovian noise processes the above would require a Lindblad master equation instead.) Contrary to previous studies, ``two-way" noise is used that affects both forward and backward evolution, to keep closer to experiment where noise cannot be turned off. This increases the fidelity decay rate at given $\sigma$ by a factor of two.

\begin{figure}
	\centering
	\includegraphics[width=80mm]{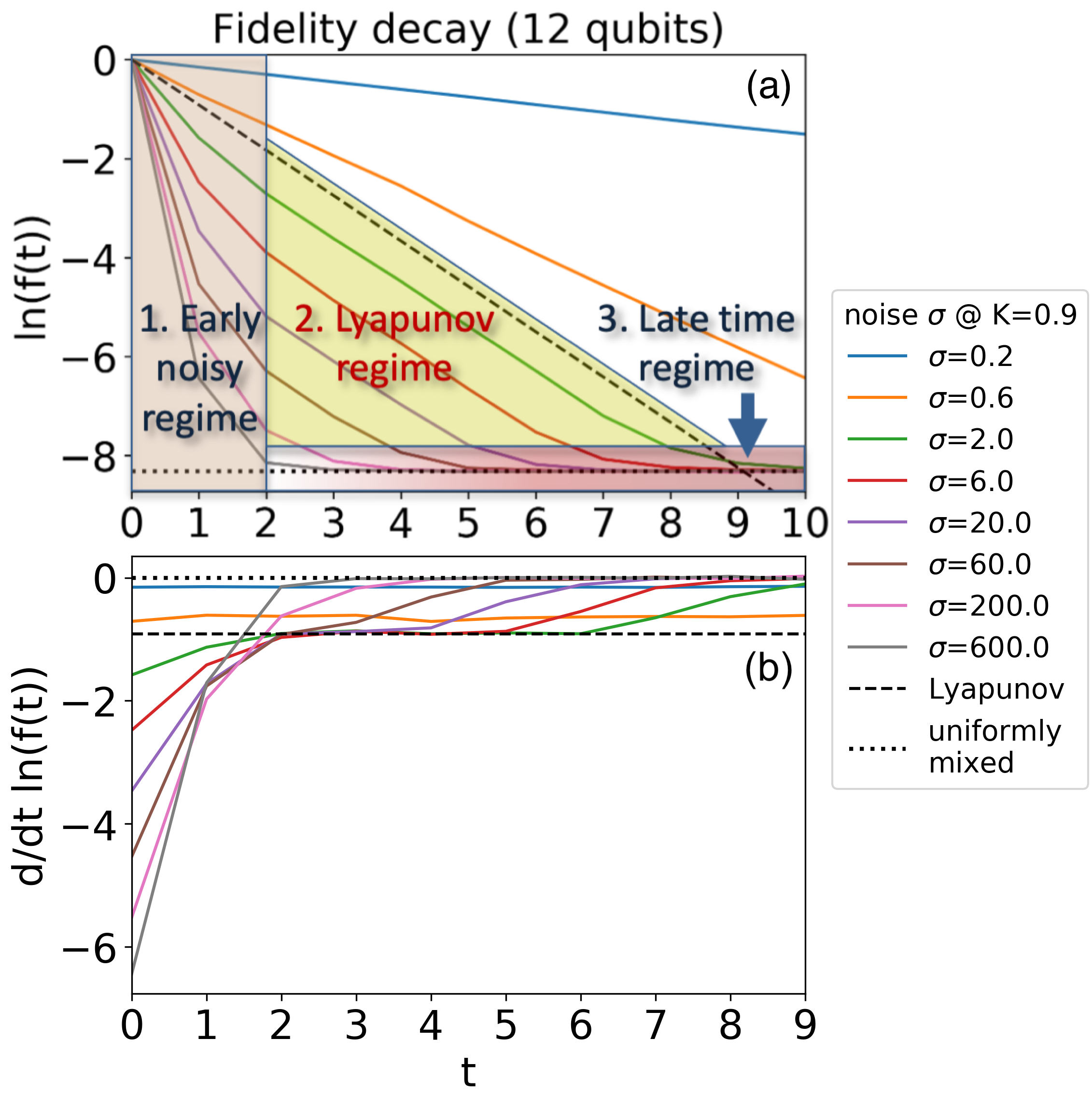}
	\caption{(a) Simulated fidelity decay $f(t)$ and (b) exponential rate of fidelity decay $-\gamma(t) \equiv d/dt \ln(f(t)) = \ln(f(t+1))-\ln(f(t))$ as noise $\sigma$ is varied. Parameters $n=12, K=0.9, L=1$ for diffusive dynamics. Dashed lines show the classical Lyapunov decay $\exp(-\lambda t)$ and dotted lines show the uniformly mixed limit $1/N$. Fidelity is averaged over 100 initial conditions and 100 realizations of the noise.}
	\label{fig:sims_fid_decay_and_slope_dif_n=10}
\end{figure}

\subsubsection{Quantum Effects of Noise} \label{sec:noise_quantum}

In Fig.~\ref{fig:sims_fid_decay_and_slope_dif_n=10} a python simulation of the QSM is implemented with random parameter noise. Over a range of noise magnitude $\sigma$ there emerge three time regimes: a fast early time decay, a slower intermediate time decay, and an even slower late time decay.  The intermediate time regime either corresponds to the FGR decay rate for sufficiently weak noise or to the Lyapunov decay rate for sufficiently strong noise.

The early and intermediate time regimes can be understood from a simple model. Derived in \cite{jalabert2001environment} but clarified in \cite{jacquod2001golden,cucchietti2004universality,cucchietti2004loschmidt} is the chaotic fidelity decay rate
\begin{equation}
	f(t) = \bar{A}(t) \exp(-\lambda t) + B \exp(- \Gamma t) + 1/N
	\label{eq:fid_jacquod}
\end{equation}
for FGR decay $\Gamma$, classical Lyapunov exponent $\lambda$, parameter-dependent $\bar{A}(t)=A'/t$, and constant $B$. The time dependence of $\bar{A}(t)$ is a simple interpolation to early time ballistic dynamics that can often be neglected once Lyapunov rate decay begins.
However it has a noticeable intermediate-time effect in some systems, such as the smooth stadium billiard \cite{cucchietti2004loschmidt}. In the QSM with $n \leq 12$ (such as Fig.~\ref{fig:sims_fid_decay_and_slope_dif_n=10}) the time dependence of $\bar{A}(t)$ for $t \ge 2$ is not noticeable, so is neglected. At intermediate times Eq.~\ref{eq:fid_jacquod} translates to $f(t) \propto \exp(-\min(\lambda, \Gamma) t)$ as the faster decay quickly depletes itself and leaves the slower decay to dominate. The distinction between early time and intermediate time regimes is most distinct in Fig.~\ref{fig:sims_fid_decay_and_slope_dif_n=10}(b) which shows the decay rate at large (but not too large) noise beginning faster than the Lyapunov rate, reducing to the Lyapunov rate for several steps, and finally dropping to zero during late time decay.

\begin{figure} [!ht]
	\centering
	\includegraphics[width=80mm]{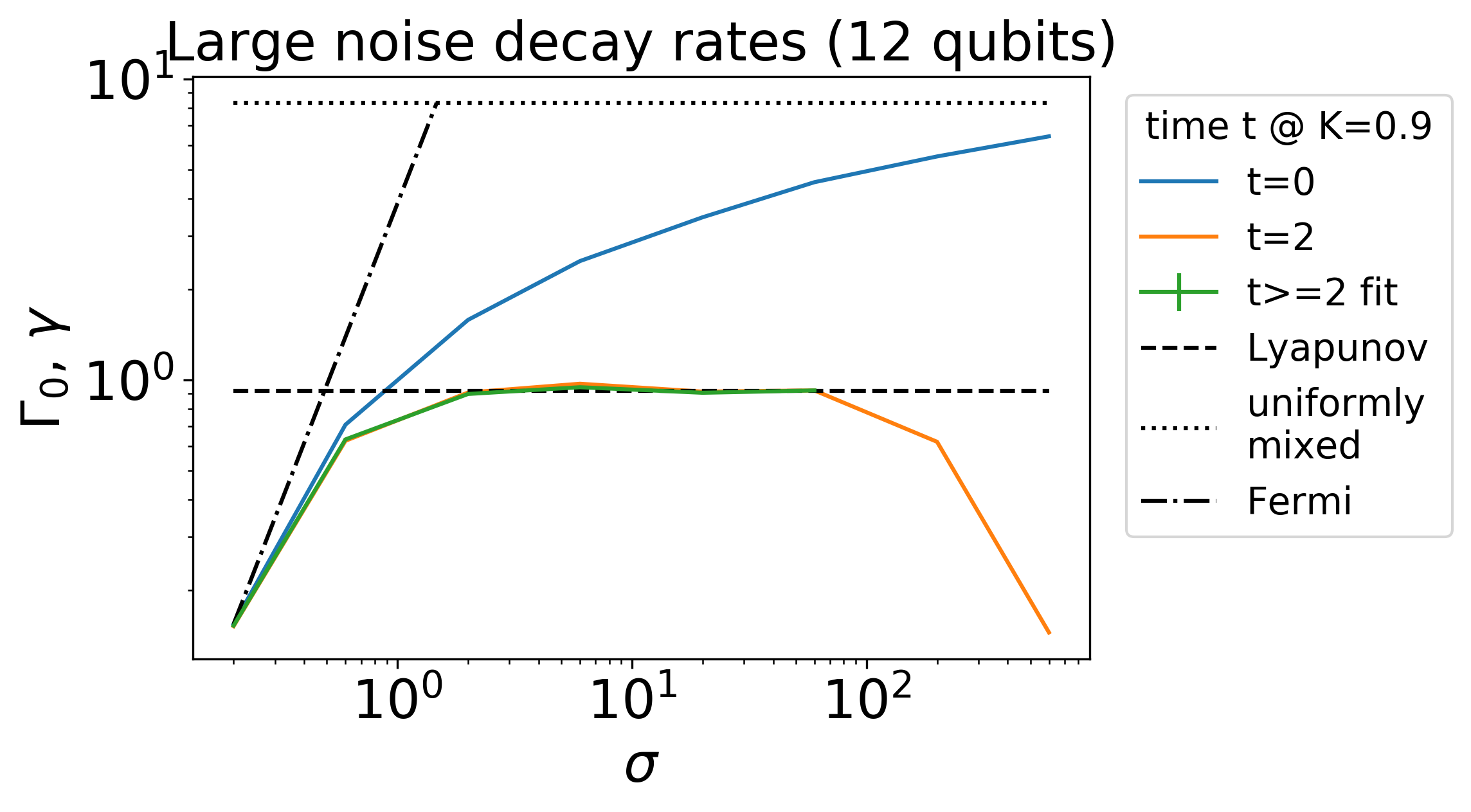}
	\caption{Simulated fidelity decay rates as functions of noise $\sigma$: initial decay rate $\Gamma_0 = -(\ln(f(1))-\ln(f(0)))$ and intermediate time decay rate $\gamma=\min(\lambda, \Gamma)$ measured at $t=2$ and measured by an exponential fit of the intermediate time regime. Same parameters as Fig.~\ref{fig:sims_fid_decay_and_slope_dif_n=10}. Fermi golden rule decay $C \sigma^2$ projected using $C$ from the smallest $\sigma$. Fidelity is averaged over 100 initial conditions and 100 realizations of the noise.}
	\label{fig:sims_fid_slope_vs_sig_dif_n=10}
\end{figure}

Late time decay slows as it asymptotically approaches the uniformly mixed limit $1/N$. The slowing can be removed by assuming the form 
\begin{eqnarray}
	f(t) = f'(t) (N-1)/N + 1/N
\end{eqnarray}
and ``unfolding" the underlying trend $f'(t)$ out of the late time decay that is dominated by the $1/N$ term. To do so just invert the relationship to get $f'(t) = N/(N-1) * (f(t) - 1/N)$. This may restore useful data points of Lyapunov decay that would otherwise be lost, as shown in Fig.~\ref{fig:lyap_decay_n=5,6}.

In Fig.~\ref{fig:sims_fid_slope_vs_sig_dif_n=10} the early and intermediate time decay rates are shown as functions of the noise over orders of magnitude. Initial decay rate $\Gamma_0$ is measured as the decay rate at $t=0$, while intermediate time decay rate $\gamma$ is measured both as the rate at $t=2$ and from fitting ($\gamma$, $t_0$) to the form $f(t) = \exp(-\gamma (t-t_0))$ for only points in the intermediate time regime, where $t \ge 2$ and $f>2/N$. The two methods agree, except at large noise where the intermediate time regime includes less than two data points so the fit becomes underconstrained. Initial decay is seen to quickly diverge from FGR, which predicts $\Gamma_0=C \sigma^2$ for noise $\sigma$ and some constant $C$. Instead it asymptotically approaches the uniformly mixed limit of reaching $f(1)=1/N$ after a single step. This slowing relative to golden rule decay is crucial to observing the Lyapunov rate on small systems. Intermediate time decay follows $\gamma=\min(\lambda, \Gamma)$ until sufficiently large noise destroys this regime.


It should be noted that the early time regime of the QSM described here has not been observed in other models or analyses \cite{jacquod2009decoherence}, though it appears to be discernible in the QSM in Ref.~\cite{benenti2004quantum}. Perturbation theory predicts a Gaussian early time decay instead \cite{jacquod2009decoherence}, but for this map discrete time steps and fast decay cannot resolve its very short time scale. The nearly constant duration $t_\text{early} \approx 2$ of this early time regime does not match predictions either. Ref.~\cite{cucchietti2004loschmidt} provides $t \ge 1/\Gamma$ as the minimal time before Lyapunov decay appears, yet for $1 \le \Gamma \le 4$ considered in Fig.~\ref{fig:sims_fid_decay_and_slope_dif_n=10} that would predict one discrete time step rather than the observed two.

Lastly, we were unable to observe the bandwidth limit mentioned in Ref.~\cite{jacquod2009decoherence} after a search up to 14 qubits. The bandwidth limit is the maximum fidelity decay rate that is reached when the perturbed Floquet eigenstates have maximally spread across the unperturbed Floquet eigenstates, as measured by the local spectral density of states. Further study may find whether this lack of a bandwidth limit is due to the initial conditions, system choice, or otherwise. Regardless it is convenient for exploring other dynamics to not have such a restriction.

\section{Lyapunov rate decay} \label{sec:lyap_bounds}

An important objective for the simulation of chaotic systems on few-qubit and NISQ devices is the experimental observation of the Lyapunov exponent. A simple method for observing the classical Lyapunov exponent in a strongly quantum system is through its fidelity decay rate, as described in Sec.~\ref{sec:noise_quantum}. (Or through related quantities like out-of-time-ordered correlators (OTOCs) discussed in Sec.~\ref{sec:conclusion}.) Here are described three limitations to observing the Lyapunov exponent in the fidelity decay rate in any chaotic quantum system, with quantitative bounds given from those limitations for the QSM.

\begin{figure}[!ht]
	\centering
	\includegraphics[width=80mm]{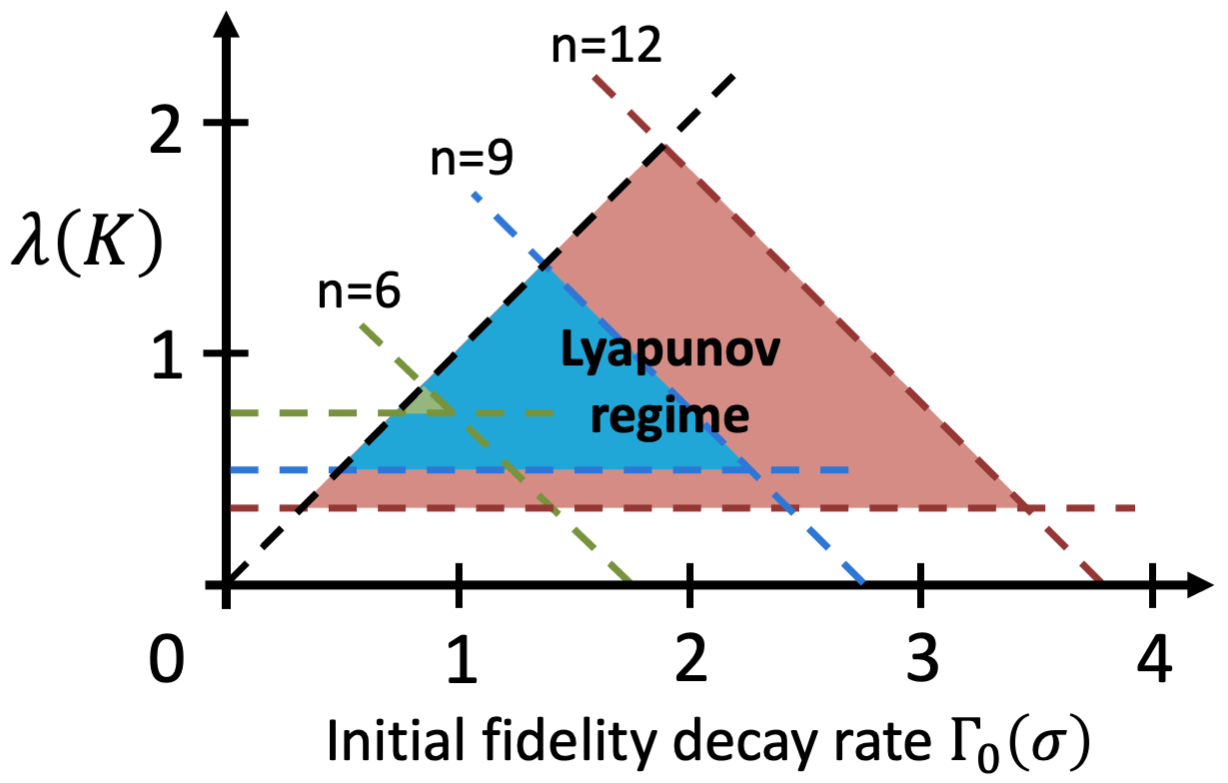}
	\caption{Parameter space showing when the Lyapunov exponent can be observed in the fidelity decay of the noisy QSM. Axes are the initial fidelity decay rate $\Gamma_0(\sigma) = -(\ln(f(1))-\ln(f(0)))$, dependent on noise magnitude $\sigma$ at given $n$, and the classical Lyapunov exponent $\lambda(\Kparam)$. Dotted lines show bounds, with solid color regions indicating when Lyapunov decay is clearly observable at different qubit numbers $n$: green for $n=6$, blue for $n=9$, and red for $n=12$.}
	\label{fig:lyap_diagram}
\end{figure}

\subsection{Three bounds} \label{sec:three_bounds}

The first limitation is that intermediate time fidelity decay of a classically-chaotic quantum system goes as $f(t) \propto \exp(-t \min(\lambda, \Gamma))$ for classical Lyapunov exponent $\lambda$ and FGR decay rate $\Gamma$. This has been well established since the first papers on Lyapunov quantum fidelity decay \cite{jacquod2001golden, jacquod2009decoherence} and is shown in the QSM in Sec.~\ref{sec:noise_quantum}. The condition for observing the Lyapunov rate is then \textbf{bound (1)}:
\begin{equation}
	\Gamma(\sigma)>\lambda(\Kparam).
\end{equation}
The initial golden rule decay rate $\Gamma_0$ and intermediate time golden rule decay rate $\Gamma$ are typically equal (see Fig.~\ref{fig:sims_fid_decay_and_slope_dif_n=10}(a) for small $\sigma$) so it is assumed that $\Gamma=\Gamma_0$ when drawing this bound in Fig.~\ref{fig:lyap_diagram}.

The second limitation is due to dynamical localization, discussed in Sec.~\ref{sec:localization}. In the weak chaos limit of small $\kparam$ (implying small $\Kparam$ and $\lambda$ for constant $\Lparam$), the localization puts a halt to chaotic diffusion. Level statistics show that as localization length $\ell \rightarrow 0$ the dynamics transition gradually from quantum chaotic to entirely regular \cite{manos2013dynamical}. This causes a transition from exponential to algebraic fidelity decay \cite{jacquod2009decoherence}. Simulations of the QSM with a simple noise model show this transition on at least five qubits, though experimentally the transition has been observed on three qubits \cite{porter2021slowed}.

\begin{figure}
	\centering
	\includegraphics[width=60mm]{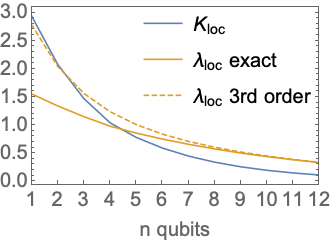}
	\caption{Parameter values below which localization occurs in the QSM, from Eq.~\ref{eq:loc_condition_K}, Eq.~\ref{eq:lyapunov}, and third order of Eq.~\ref{eq:bound_2}, respectively.}
	\label{fig:exact_lyap_values}
\end{figure}

One can obtain a diffusion bound specific to when Lyapunov decay is desired by using small $\Lparam$ and large $N$, which increase the chance of diffusion and therefore Lyapunov decay.
From Eq.~\ref{eq:loc_condition_K} these limits enter the $0<\Kparam_\text{loc}<1$ cantori regime, meaning diffusion occurs when $\Kparam > \Kparam_\text{loc} \approx 3.12 \Lparam^{4/5} N^{-2/5}$. Assuming $N \gg 17.3 \Lparam^2$ implies that $0<\Kparam_\text{loc} \ll 1$, allowing use of the power expansion of $\lambda(\Kparam_\text{loc})$ in Eq.~\ref{eq:lyapunov} to derive \textbf{bound (2)}:
\begin{flalign}
	\lambda(\Kparam) &\gtrsim \lambda(\Kparam_\text{loc}) \nonumber \\
	 &\approx 1.77 \frac{\Lparam^{2/5}}{N^{1/5}} - 0.23 \frac{L^{6/5}}{N^{3/5}} + 2.43 \frac{L^{8/5}}{N^{4/5}} + ...
	\label{eq:bound_2}
\end{flalign}
This bound is approximate for several reasons. First, the diffusion bound is heuristically derived and imprecise. 
Second, since localization occurs after about $\ell$ time steps and $\ell$ is large near the bound, Lyapunov decay could still occur during the initial classical diffusion, suggesting this bound is too strict. However in practice distinguishing the transition to localization may be difficult in few-qubit experiments. The algebraic decay due to localization appears to pass through many ``exponential" rates  and may be mistaken for a single step of exponential Lyapunov decay during the initial classical diffusion. This suggests two steps of Lyapunov decay are needed, similar to Figs.~\ref{fig:sims_fid_decay_and_slope_dif_n=10} with $\sigma=20$, which will be accounted for in the bound (3).
Third and lastly, it was assumed that $n = \log_2(N) \gg 4$ (when $\Lparam=1$). For the small $n$ of current experiments one should instead calculate $K_\text{loc}$ from Eq.~\ref{eq:loc_condition_K} and plug into the exact $\lambda(\Kparam)$ of Eq.~\ref{eq:lyapunov} directly. This approach is compared to the third order expansion in Fig.~\ref{fig:exact_lyap_values}.

The third limitation is competition from the early time and late time decays. Sufficient noise leading to short decay can cause the intermediate time Lyapunov decay to get squeezed out. Figs.~\ref{fig:sims_fid_decay_and_slope_dif_n=10} and \ref{fig:sims_fid_slope_vs_sig_dif_n=10} demonstrate this at large noise where the Lyapunov rate is no longer visible. Since early time decay is numerically observed to end at $t=t_\text{early} \approx 2$ fairly consistently and late time decay is seen to begin at $f=a_\text{late}/N$ with $a_\text{late} \approx 2$, one can write a bound on the initial decay rate $\Gamma_0$ to ensure these two don't meet as $\exp(-\Gamma_0 t_\text{early}) \gtrsim a_\text{late}/N$. A stricter bound requiring a clear Lyapunov signature must include a number of intermediate time steps $t_\text{lyap}$ at the Lyapunov rate, to avoid the ambiguity with algebraic decay. We choose $t_\text{lyap} = 2$ as a reasonable minimal number of Lyapunov steps. This gives $\exp(-\Gamma_0 t_\text{early} - t_\text{lyap} \lambda) \gtrsim a_\text{late}/N$ or \textbf{bound (3)}:
\begin{equation}
	\Gamma_0(\sigma) \lesssim (\ln(N/a_\text{late}) - t_\text{lyap} \lambda(\Kparam)) / t_\text{early}.
	\label{eq:bound_3}
\end{equation}
Unfolding the late time regime may be helpful here by reducing $a_\text{late}$.
Since $\Gamma_0$ varies with $\sigma$ more slowly than the quadratic FGR prediction, as shown in Fig.~\ref{fig:sims_fid_slope_vs_sig_dif_n=10}, so Eq.~\ref{eq:bound_3} can be satisfied at surprisingly large noise $\sigma$. 

For very large systems, experimental statistics become limited by the achievable number of samples $S$. In that case the late time limit changes by $a_\text{late}/N \rightarrow 1/\sqrt{S}$.

These three bounds taken together form a triangular ``Lyapunov regime" in parameter space as shown in Fig.~\ref{fig:lyap_diagram}.

\subsection{Qubit and noise requirements} \label{sec:requirements}

\subsubsection{Qubit minimum}
\label{sec:qubit_min}

The first conclusion from the three bounds in Fig.~\ref{fig:lyap_diagram} is that a minimum of six qubits is needed for a Lyapunov regime of non-zero area in parameter space to exist. This is a theoretical lower bound for any platform. It assumes purely unitary noise that at small noise strength produces FGR fidelity decay.

If incoherent noise is also present, as is common in present day devices, the lower bound on qubits only becomes more strict. In Ref.~\cite{porter2021slowed} it was shown in simulation that incoherent Lindblad noise cannot alone generate fidelity decay at the Lyapunov rate, but when paired with unitary noise can produce a decay rate that is the sum of the Lyapunov rate $\lambda$ and Lindblad rate $\nu_\text{eff}$, assuming the two noise processes are independent. This acts only to tighten bound (3) by increasing the total decay rate from $\Gamma_0$ to $\Gamma_0 + \nu_\text{eff}$, shortening the simulation time. Therefore the minimum number of qubits needed grows with the size of the incoherent noise. Dependence on the unitary noise will be discussed in the next section.

\begin{figure}[!ht]
	\centering
	\includegraphics[width=80mm]{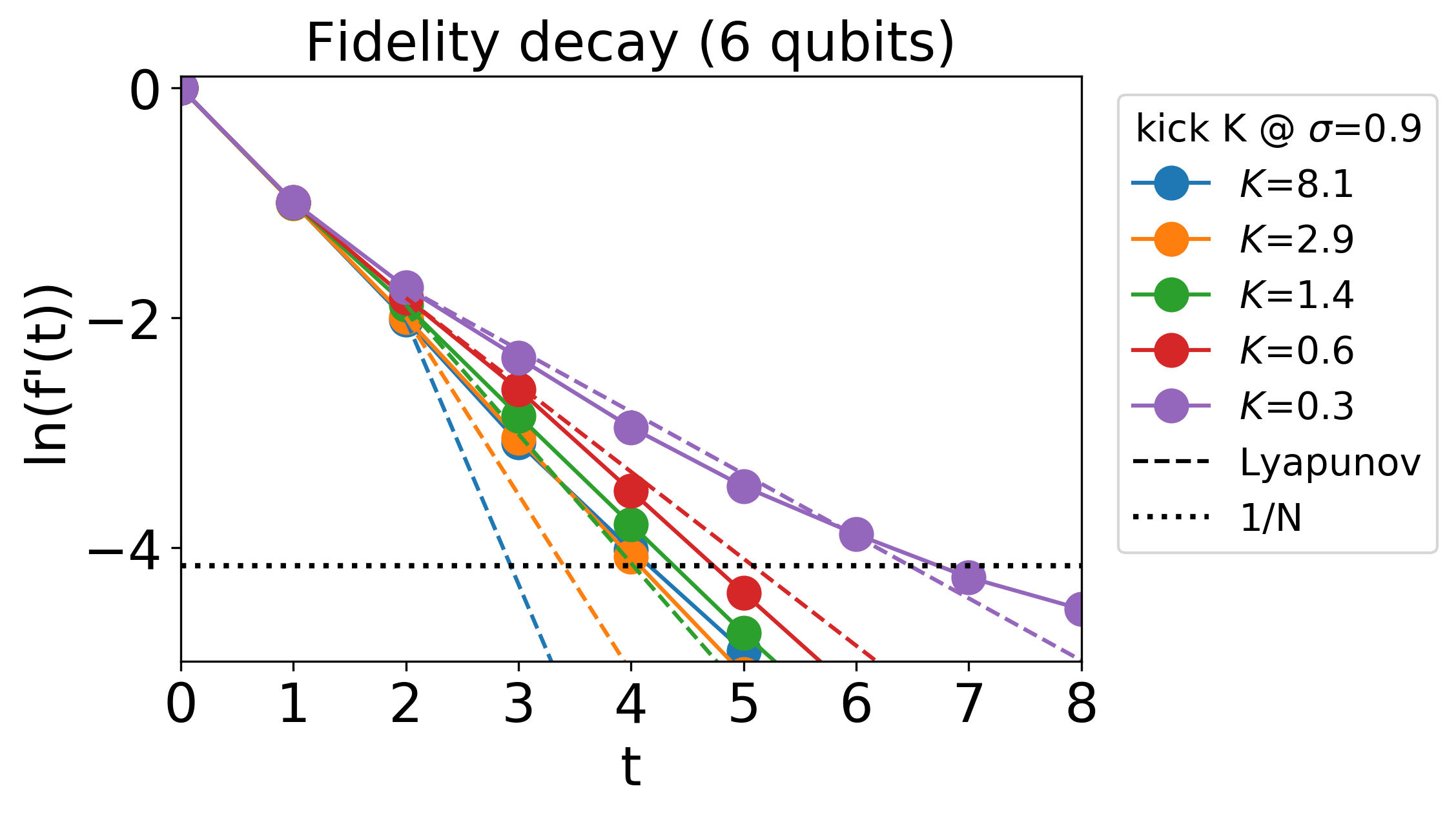}
	\caption{Simulations of minimal-qubit Lyapunov decay, where $n=6, \sigma=0.9, L=1$ and $\Kparam$ is varied. Dashed lines show theoretical Lyapunov decay projected from a start at $t_\text{early}=2$. Comparing the golden rule decay for $\Kparam \ge 2.9$ to Lyapunov decay at $\Kparam=0.6$ shows a small but noticeable difference. Localization-slowed decay at $\Kparam=0.3$ has a subexponential trend that is difficult to clearly distinguish from Lyapunov decay at intermediate times. Localization occurs for $\Kparam<\Kparam_\text{loc}=0.59$. Fidelity is averaged over 62 initial conditions and 1000 realizations of the noise.}
	\label{fig:lyap_decay_n=5,6}
\end{figure}

To verify the ability to observe Lyapunov-limited decay on six qubits in practice, a simulation with the stochastic unitary noise model of Sec.~\ref{sec:noise_models} is provided in Fig.~\ref{fig:lyap_decay_n=5,6}. In an experiment the noise cannot be easily varied, except to artificially increase it (which may be valuable in this context), but $\Kparam$ can be varied freely. For a noise magnitude within the appropriate triangle in Fig.~\ref{fig:lyap_diagram}, scanning $\Kparam$ reveals the transition in the decay rate from FGR to the Lyapunov rate $\lambda(\Kparam)$ for times $t > t_\text{early} \approx 2$. However this ends at the transition to localization at $\Kparam_\text{loc}$ at the bottom of the triangle. Two time steps of Lyapunov rate decay are needed to distinguish it from the localized algebraic rate decay. For six qubits the Lyapunov regime ranges for noise and $\Kparam$ are both minuscule, so the effect is very subtle and the second time step is slightly faster than the Lyapunov rate. This is despite unfolding the late time behavior per Sec.~\ref{sec:effects_of_noise}, which should slightly increase the number of Lyapunov steps observed. Still, the numerical results conform well to the theoretical prediction that six qubits is the minimal system size at which a Lyapunov decay rate may barely be observed.

\subsubsection{Noise scaling factors}

One can determine the maximum allowable initial fidelity decay rate due to unitary noise within the Lyapunov regime for a given number of qubits $n$, which we call $\Gamma_{0,\text{max}}(n)$, from Fig.~\ref{fig:lyap_diagram}. Meanwhile experimental data from Ref.~\cite{porter2021slowed} shows that executing the QSM on IBM-Q's platform with $n \ge 4$ produces an initial decay too fast to resolve the rate $\Gamma(t=0, n) \equiv \Gamma_0(n)$ from the $1/N$ term in Eq.~\ref{eq:fid_jacquod}. Instead experiments with $n=3$ provide a baseline at the current three-qubit decay rate $\Gamma_0(n=3)$. By comparing $\Gamma_0(3)$ to the target $\Gamma_{0,\text{max}}(n)$ at given $n$ while accounting for the scaling with $n$, one can determine the factor of error reduction needed to reach the Lyapunov regime.

From a hardware perspective it is more useful to focus on the reduction in error per \code{CNOT} gate, neglecting the small contribution from single-qubit gate error. To do so, note the simple relationship at given $n$ between the first-step unfolded decay rate $\Gamma_0$, the error per \code{CNOT} gate $\epsilon$, and the ``effective" gate depth $\Gparam$, which is the gate count divided by the error improvement due to gate parallelization \cite{porter2021slowed}. These are related by
\begin{flalign}
	\exp(-\Gamma_0(n)) &= (1-\epsilon(n))^{\Gparam(n)} \\
	\epsilon(n) &= 1 - \exp(-\Gamma_0(n)/\Gparam(n) ) \nonumber \\
	&\approx \Gamma_0(n)/\Gparam(n) \quad \text{ for } \Gparam(n) \gg \Gamma_0(n).
	\label{eq:error_decay_relation}
\end{flalign}

The goal is to calculate the required gate error reduction 
\begin{flalign}
	\label{eq:r_eG_G0max}
	r(n) &\equiv \epsilon(n) / \epsilon_\text{max}(n) \nonumber\\
	&= \epsilon(n) * \Gparam(n) / \Gamma_{0,\text{max}}(n)
\end{flalign}
from the present day error $\epsilon(n)$ (based on error scaling models and recent benchmarks) to the maximum allowable error $\epsilon_\text{max}(n)$ (based on gate scaling and our three bounds) in order to reach the Lyapunov regime with $n$ qubits. We consider three dominant factors in this analysis: $\Gamma_{0,\text{max}}(n)$ increasing with $n$ per Fig.~\ref{fig:lyap_diagram}; $G(n)$ increasing with $n$ as the circuit decomposition grows in size per Ref.~\cite{porter2021slowed}; and $\epsilon(n)$ increasing with $n$ due to crosstalk from additional qubits. 

The value of $\Gamma_{0,\text{max}}(n)$ comes from the intersection of bounds (2) and (3) at the right corners of each triangle in Fig.~\ref{fig:lyap_diagram}. Setting both bounds to equalities yields
\begin{flalign}
	\label{eq:a_scaling}
	\Gamma_{0,\text{max}}(n \ge 6) \approx& \ln(N/2)/2 - 1.77 \frac{\Lparam^{2/5}}{N^{1/5}} \nonumber \\
	&+ 0.23 \frac{L^{6/5}}{N^{3/5}} - 2.43 \frac{L^{8/5}}{N^{4/5}} \\
	\approx&\, n \ln(2)/2 \quad \text{ for large n} \nonumber
\end{flalign}
using third order in $\lambda_\text{loc}$ as a decent approximation at large $n$ according to Fig.~\ref{fig:exact_lyap_values}. $\Gamma_{0,\text{max}}(n<6)$ is not defined as the Lyapunov regime occupies zero area in parameter space. We will keep $\Lparam=1$ to make bound (2) minimally strict.


That leaves the scaling of $\Gparam(n)$ and $\epsilon(n)$, which depend strongly on the architecture of the quantum platform. To allow for variance in the architectures of future platforms, we will consider a range of possible scalings for each of these.

\subsubsection{Architecture dependent scaling}

The scaling of effective gate depth $\Gparam(n)$ and average two-qubit gate error $\epsilon(n)$ with increasing number of qubits $n$ depend primarily on four features: the connectivity of qubits; the ability to reduce the two-qubit gate count or depth using circuit optimization algorithms; the ability and effectiveness of parallelizing gates to reduce total error; and the model of crosstalk. 

The gate count grows when poor connectivity requires additional \code{SWAP} gates to communicate between unconnected qubits, but shrinks if circuit optimization algorithms can be efficiently employed. The effective gate depth $G(n)$ then depends on gate count, the ability to parallelize an algorithm's gates, and the error reduction due to that parallelization on the given hardware.

Meanwhile the two-qubit gate error $\epsilon(n)$ scales with crosstalk, but must account for base error without crosstalk and error enhancement during complex dynamics \cite{porter2021slowed}. A simple model of these effects on error would be 
\begin{flalign}
	\label{eq:error_scaling_model}
	\epsilon(n) = a_\text{dyn}(\epsilon_\text{base} + \epsilon_\text{cross} n_\text{nb,act} \nonumber \\ 
	+ \epsilon_\text{cross,inact} n_\text{nb,inact})
\end{flalign}
where the base error $\epsilon_\text{base}$ is increased by crosstalk proportionally to the number of neighbor qubits and then scaled by the dynamics. Dynamical decoupling experiments have shown that even inactive ground state qubits can contribute to crosstalk \cite{tripathi2021suppression}. Since inactive and active qubits may contribute different magnitudes of noise, we count them separately as $n_\text{nb,inact}$ and $n_\text{nb,act}$ with associated crosstalk errors. Dynamics may affect all error terms differently, or even have its own dependence on $n$. However, for simplicity we assume the QSM dynamics contributes a constant, uniform factor $a_\text{dyn}$.

With this model, the effects of architecture on the scalability of the Lyapunov regime can be explored with minimal assumptions by considering the extreme worst and best case scenarios. For specific platforms these scenarios can then be tailored to a more appropriate range.

The hypothetical worst case architecture has: a linear qubit topology to produce minimal connectivity; ineffective circuit optimization algorithms; either the inability to parallelize or no total error reduction from parallelization; and the maximum crosstalk errors $\epsilon_\text{cross}, \epsilon_\text{cross,inact}$ that fit the benchmarking data and architecture. Ineffective circuit optimization could be due to the NP-completeness of finding an optimal circuit decomposition \cite{botea2018complexity}. The number of extraneous connections to inactive qubits away from the chain ends is assumed negligible, as otherwise they would be included in the simulation for increased connectivity and better overall performance.

The hypothetical best case architecture we will consider has (without error correction): all-to-all connectivity; an optimal gate count achieved by either the algorithm or an efficient circuit optimization algorithm; full parallelization of two-qubit gates throughout the algorithm; the smallest crosstalk errors that fit the benchmarking data and architecture; and no larger device than necessary, $n = n_\text{device}$.

Effective gate depth $\Gparam(n)$ of the QSM combines the exact algorithm in Refs.~\cite{benenti2001efficient, porter2021slowed} with factors for poor connectivity in the worst case or effective parallelization and circuit optimization in the best case. The algorithmic gate count for a forward-and-back simulation is $4n^2 - 4n$ \code{CPHASE} gates. Linear topology without optimization requires chains of \code{SWAP} gates to connect qubits an average of $(n+1)/3$ connections away, with two \code{SWAP}s needed per connection beyond the first. This incurs a cost of approximately $2((n+1)/3 -1)$ \code{SWAP}s per \code{CPHASE} gate on average. If \code{SWAP} and \code{CPHASE} gates are converted to native \code{CNOT} or \code{MS} gates, at a rate of three gates per \code{SWAP} and two per \code{CPHASE}, then the factor increase of native gates due to linear topology is $(n-1)$. Parallelization can typically reduce gate depth by up to $n/2$, while circuit optimization depends on both topology and the software package being used. For simplicity, here we assume a factor of 2 improvement in gate count for all-to-all connectivity that might be achievable in the future based on the $80\%$ reduction we observed on three qubits using Qiskit.

\begin{table*}[!ht]
	\centering
	\begin{tabular}{| p{36mm} | p{24mm} | p{30mm} | p{30mm} | }  
		\hline
		Bounds & $\epsilon(n)$ (approx.) & $\Gparam(n)$ (approx.) & $r(n)$ (approx.) \\
		\hline
		IBM-Q worst & $0.05$ & $8n(n-1)^2$ & $0.9n^2 + 2n$ \\
		\hline
		IBM-Q best & $0.008 n + 0.004$ & $8(n-1)$ & $0.1n + 0.5$ \\
		\hline
		IonQ worst & $0.3$ & $8n(n-1)$ & $6n + 20$ \\
		\hline
		IonQ best & $0.03$ & $4(n-1)$ & $0.3 + 0.8/n$ \\
		\hline
	\end{tabular}
	\caption{ The error reduction per gate $r(n)$ needed to observe the Lyapunov regime as a function of $n$ qubits. The scaling factors of gate error $\epsilon(n)$ and effective gate depth $\Gparam(n)$ are bounded by two extreme scenarios based on uncertainties in future architecture. $r(n)$ is calculated from Eq.~\ref{eq:r_eG_G0max} to next to leading order. The factor $\Gamma_{0,\text{max}}(n)$ is approximated as $0.43n - 1.6$, a close linear fit to its exact form Eq.~\ref{eq:a_scaling} in the relevant range $6 \le n \le 12$. Error models and gate depths specific to two leading quantum hardware platforms, IBM-Q and IonQ, are considered. }
	\label{table:scaling_params}
\end{table*}

For $\epsilon(n)$, the worst and best cases on different architectures lead to different simplifications of Eq.~\ref{eq:error_scaling_model}. In general, crosstalk comes from neighbors, in both space and frequency, of the two qubits targeted by the gate. Only nearest neighbors in space will be considered here.

\begin{figure}[!t]
	\centering
	\includegraphics[width=80mm]{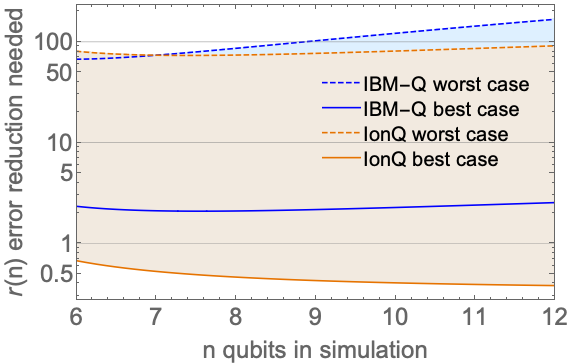}
	\caption{The factor $r(n)$ (see Table~\ref{table:scaling_params}) by which error per two-qubit gate must be reduced on IBM-Q and IonQ for them to reach the Lyapunov regime, as a function of number of qubits $n$ used in the simulation. For each, two possible future cases are considered: the best case assumes all-to-all connectivity and maximal effective parallelization, while the worst case makes the fewest such assumptions based on current hardware. The exact expression for $\Gamma_{0,\text{max}}(n)$ is used.}
	\label{fig:abc_scaling}
\end{figure}

\subsubsection{IBM-Q scaling}

For the open access IBM-Q platform, the error model Eq.~\ref{eq:error_scaling_model} can be roughly tailored based on published results.

The crosstalk error from inactive ground state qubits is difficult to disentangle given our limited data and the number of other unknowns. Ref.~\cite{tripathi2021suppression} did show that on IBM-Q inactive neighbor qubits in various states can cause fidelity oscillations in an active qubit, increasing average error. This could influence RB and other experiments. Fortunately, based on the best and worst case topologies, the number of inactive neighbor qubits is negligible for the target simulation sizes of $n \ge 6$. This is due to the fact that linear topology only has neighbors at the ends of the chain and the assumption that the best case topology is no larger than necessary. When instead using data from smaller systems to fit parameters, we can reasonably attribute all error to other sources without compromising the goal of estimating best and worst case scenarios. Therefore, inactive qubit crosstalk can be excluded from the present analysis by setting $\epsilon_\text{cross,inact} \approx 0$.

The remaining parameters can be fit in the best and worst cases to several benchmarking experiments. Without inactive qubit crosstalk, one can use IBM-Q's reported RB gate error from Ref.~\cite{porter2021slowed} as the base error, $\epsilon_\text{base} \approx 0.006$. For dynamics and crosstalk it is useful to compare our $n=3$ QSM experiment in that reference to $n=2$ RB for the same device and time. The QSM gate error of $\epsilon \approx 0.03$ was about $5\times$ larger.

In the worst case, this increase in error might be attributed entirely to crosstalk from the single active neighbor, with no increase due to dynamics. This produces a model $\epsilon(n) = 0.006(1 + 4 n_\text{nb,act})$. However, since the worst case is assumed to have linear topology, its number of active neighbors approaches $n_\text{nb,act} = 2$ for the target system sizes of $n \ge 6$. So the error model for those system sizes simplifies to the constant $\epsilon(n) = 0.05$.

In the best case, the increase in error can be attributed mostly to the constant dynamics term, $a_\text{dyn}$, with a bare minimum of crosstalk. This minimum crosstalk comes from another experiment Ref.~\cite{mckay2019three} that demonstrates the presence of active qubit crosstalk on IBM-Q devices. That experiment finds a wide range of increased gate error as a neighbor qubit is switched from inactive to active. After attributing inactive qubit crosstalk to base error as discussed above, and excluding one outlier where crosstalk greatly decreased the error, the smallest change out of five experiments is a factor of $1.4$ increase in error. This suggests $\epsilon_\text{cross} \approx 0.4 \epsilon_\text{base}$. Assuming this minimum crosstalk, the QSM experiment implies $a_\text{dyn} \approx 4$. Since the best case is assumed to have all-to-all connectivity and $n = n_\text{device}$, each two-qubit gate has many active neighbors $n_\text{nb,act} = n-2$. The total error model of the QSM is then $\epsilon(n) = 0.004 + 0.008 n$.

For effective gate depth $\Gparam(n)$, \code{CPHASE} and \code{SWAP} gates are converted to \code{CNOT} gates. In the worst case, the algorithmic gate count of $8n(n-1)$ is modified by a linear topology factor of exactly $(n-1)$ to connect distant qubits, for a total of $\Gparam(n) = 8n(n-1)^2$. In the best case, factors of $2/n$ for parallelization and $1/2$ for optimization, assumed to be constant with $n$, reduce the total to $\Gparam(n) \approx 8(n-1)$.

Inserting these results and the derived scaling of $\Gamma_{0,\text{max}}(n)$ in Eq.~\ref{eq:a_scaling} into Eq.~\ref{eq:r_eG_G0max} results in Table~\ref{table:scaling_params} and the blue lines in Fig.~\ref{fig:abc_scaling}. The figure uses the exact expression for $\Gamma_{0,\text{max}}(n)$ but otherwise reflects Table~\ref{table:scaling_params}.

Using these, one can estimate the current initial decay rate on six qubits, assuming existing linear topology. The tradeoff of $\epsilon_\text{cross}$ and $a_\text{dyn}$ between worst and best cases is still applicable. Then approximating $n_\text{nb,act} \approx 2$, the range of error is $\epsilon(6) \in [0.04, 0.05]$. For $\Gparam(n)$ the best case now uses linear topology but with improvements from highly effective parallelization and circuit optimization, for a form $\Gparam(n) \approx 8(n-1)^2$. The range is then $\Gparam(6) \in [200, 1200]$. Combining these yields $\Gamma_0(6) = \epsilon(6) \Gparam(6) \in [8, 60]$, far from the desired $\Gamma_{0,\text{max}}(6) \approx 0.9$. This shows that the decay rate is too fast for even a single useful time step.

The forms of control IBM-Q could exert within the range in Fig.~\ref{fig:abc_scaling}, aside from error reduction, are to increase qubit connectivity, to scale up their circuit optimization algorithms, and to increase coherence times of active qubits towards the times of idle qubits to make parallelization more effective. Of these, connectivity would provide the clearest gain by both decreasing gate count and increasing ability to parallelize, at the (smaller) cost of additional crosstalk. 

The apparent reason for poor connectivity in current IBM-Q devices is their focus on hexagonal lattices for the goal of fault tolerance \cite{IBM2020Hexagon}. However other groups have explored all-to-all connectivity in superconducting architectures  \cite{roy2020programmable,marinelli2021dynamically,mckay2019three}, though not yet at the scale of six qubits. 

If connectivity is kept low, there is still much possiblity for improvement. The \code{SWAP} gates caused by low connectivity increase the potential for optimization over the algorithmic gates alone. Achieving scalable circuit optimization and/or highly effective parallelization in a sparse topology could reduce $G(n)$ to $O(n^2)$ or even $O(n)$, equalling or even surpassing the ``best" case $r(n)$ of all-to-all connectivity. This would be particularly attractive if reducing crosstalk in a highly connected topology proves too difficult.

\subsubsection{IonQ}

Ion traps are another leading architecture type, and it is worth estimating how far they might be from the Lyapunov regime. A single \code{CNOT} gate from the QSM decomposition can be converted to a single M{\o}lmer-S{\o}rensen (MS) gate plus single-qubit gates \cite{maslov2017basic}, so the same scaling of two-qubit gates applies. IonQ is chosen here to represent current capability since they have demonstrated algorithms on an 11-qubit quantum computer \cite{wright2019benchmarking}, making them competitive with IBM-Q. They benefit from all-to-all connectivity, though they do not presently offer parallel gates. (Limited parallel gates have been demonstrated on IonQ, but suffer from large control error and gate duration \cite{grzesiak2020efficient}.) The quantum charge-coupled device (QCCD) architecture employed by Quantinuum and others is designed for large parallelization \cite{kielpinski2002architecture}, so fully parallel gates will still be considered. 

The long idle qubit coherence times of ion traps would at first glance suggest little gain from parallelization, as two gates in parallel would have similar error as two gates in series \cite{porter2021slowed}. However, ion trap two-qubit gates are physically realized as all-qubit gates, with motional dephasing and other time-sensitive control errors as major sources of error \cite{sutherland2022one}. Which errors dominate is determined by the form of calibration used \cite{beckprivate}, among other factors, so the benefit of parallelization, while real, appears highly context dependent.

Equation~\ref{eq:error_scaling_model} can again be roughly tailored for IonQ. One unique feature of their device is that, due to their calibration procedure, the main crosstalk error is entirely experienced by the neighbors in the form of Raman beam tails \cite{beckprivate}, providing a physical reason to set $\epsilon_\text{cross,inact} = 0$ for qubits that get measured. If one restricts crosstalk discussion to this beam tail type, it also implies that only spatial neighbors in a chain of ions can be involved in crosstalk, despite the capability of all-to-all gates. This means $n_\text{nb,act} \approx 2$ for the target $n \ge 6$ simulations.

More pessimistically, other effects could still cause error to increase with $n$. For one, the difficulty of constructing each two-qubit gate increases with the device size $n_\text{device}$, due mostly to tightly packed motional modes requiring longer gates to correctly cancel all motion \cite{grzesiak2020efficient, figgatt2019parallel}. This could cause gate error to scale as $\epsilon_\text{base} \sim n_\text{device}$. Another is the axial heating rate which can grow due to the axial mode frequencies decreasing with more qubits \cite{li2020generalized}. However, neither of these effects need be accounted for in the present analysis. Since we will use benchmarking data from an 11-qubit device and the present scope is 12 or fewer qubits, these considerations are accounted for in the data and should not cause further difficulties in our range of interest.

The remaining parameters can be fit from data. This will entirely rely on their benchmarking paper Ref.~\cite{wright2019benchmarking}. They first prepare a bell state as a measurement of base error, $\epsilon_\text{base} \approx 0.03$.

A range of crosstalk error $\epsilon_\text{cross}$ can be extracted from another benchmark in that reference, the 10-qubit Hidden Shift, again with simple, Clifford dynamics. While they do not report error per two-qubit gate, if one attributes the extra error solely to two-qubit gates (as seems reasonable from their consistent total error with increasing single-qubit gate count), the gate error can be calculated by comparing expected and observed fidelities in the equation $0.4/0.6 \approx ((1-\epsilon)/(1-0.03))^{10}$. This yields $\epsilon = \epsilon_\text{base} + 2 \epsilon_\text{cross} \approx 0.07$ and therefore $\epsilon_\text{cross} \approx 0.02$. An uncertainty of about $0.01$ can be estimated from the error bars. More encouraging results from IonQ's Bernstein-Vazirani experiment in the same reference suggest $\epsilon_\text{cross} \approx 0$. Together they provide a range from best to worst case of $\epsilon_\text{cross} \in [0, 0.03] = [0,1] \epsilon_\text{base}$.

The effect of QSM dynamics $a_\text{dyn}$ requires a QSM experiment to determine. In the absence of such experiments on IonQ, we borrow the range from IBM-Q of $a_\text{dyn} \in [1,4]$. Unlike IBM-Q, the uncertainties in $\epsilon_\text{cross}$ and $a_\text{dyn}$ are not correlated, so the final model has a larger spread. The range is a constant in $\epsilon(n) = [0.03, 0.3] $.

For the scaling of $\Gparam(n)$, the abilities to parallelize two-qubit gates and optimize the circuit can reduce gate depth. Starting from the worst case gate count of $\Gparam(n) = 8n(n-1)$, reducing by $1/n$ for parallelization and $1/2$ for circuit optimization yields $\Gparam(n) \approx 4(n-1)$.

These results and the associated $r(n)$ expressions are given in Table~\ref{table:scaling_params} and by the orange lines in Fig.~\ref{fig:abc_scaling}. The combination of constant scaling of gate error and high connectivity is encouraging. Parallel gates would reduce $r(n)$ by an order of magnitude, while another order of magnitude lies in the uncertain factors for crosstalk and especially dynamics, which require further study.

\subsubsection{Discussion}

The ranges of improvement needed according to Fig.~\ref{fig:abc_scaling} are sobering yet encouraging. In the absence of architectural improvements, both platforms lean towards the worst case, requiring two orders of magnitude in error reduction for this relatively simple objective. However, the best cases show the potential of improved architecture, such as the QCCD architecture that realizes parallel gates while likely maintaining constant scaling of crosstalk. Whether this parallelization causes an effective error reduction is a topic for future study.

One remaining uncertainty is in the scaling of the dynamical factor $a_\text{dyn}$ with $n$ under complex, non-Clifford dynamics. It was assumed here to be constant, and measured from a three-qubit QSM simulation, but further study is needed to investigate the dependence on $n$. This is important not only for the QSM but for any plan for quantum advantage, which all require non-Clifford gates \cite{gottesman1998heisenberg}. 

Another caveat that inflates these $r(n)$ estimates is that this analysis attributed experimentally observed fidelity decay rates entirely to the FGR initial decay rate $\Gamma_0$ caused by unitary errors. In contrast, Ref.~\cite{porter2021slowed} found that experimental results from IBM-Q were qualitatively explained by an incoherent Lindblad model but not by a unitary noise model. Large incoherent noise is observed even on idle qubits on IBM-Q, and only increases in the presence of gates. This all implies that the present day FGR rate $\Gamma_0$ is much smaller than suggested by the gate error $\epsilon$, and that the incoherent decay rate $\nu_\text{eff}$ is large and must be accounted for. The main consequence is a much stricter bound (3) that requires more than six qubits to satisfy, as discussed in Sec.~\ref{sec:qubit_min}. The analysis of the error reduction $r(n)$ still largely applies to reducing incoherent noise, with the correction to $\Gamma_{0, \text{max}}(n)$ depending only on $\nu_\text{eff}$ and the corresponding minimum number of qubits. This requires a careful experimental analysis of the relative roles of unitary and incoherent errors in a given device, which is beyond the scope of this paper.

Lastly, progress towards these reductions in error through $r(n)$ can already be made through software-level noise mitigation methods. These include: dynamical decoupling to suppress decoherence on idle qubits \cite{viola1999dynamical}, randomized compiling to reduce Markovian noise to stochastic Pauli errors with reduced worst-case error \cite{wallman2016noise}, and measurement error mitigation \cite{IBM2021measurement}. Even non-Markovian noise can be mitigated by techniques such as randomized dynamical decoupling \cite{viola2005random}. These reduce the burden on hardware, but until fault tolerance is achieved they are only a partial solution.

\section{Conclusion} \label{sec:conclusion}

In this paper the quantum simulation of the Lyapunov exponent was studied as a test for semiclassical chaos in a quantum system and minimal bounds were given for its observation on future few-qubit quantum devices. The Lyapunov exponent is observable as the slowing of quantum fidelity decay down to the classical Lyapunov rate when the dynamics are chaotic (diffusive) and noise is large yet not overwhelming. The mechanism is a competition between additive decay rates in which the slowest rate lasts the longest and becomes observable.

To find the most plausible route to this Lyapunov regime, a particularly resource-efficient quantum map was chosen, the quantum sawtooth map. The classical and quantum dynamics of this map were described, bounds on its dynamical localization were given, and simulations of its time and noise regimes were shown. The specifics of the quantum sawtooth map were then used to derive three quantitative bounds that must be satisfied to observe Lyapunov rate fidelity decay. Together those bounds formed a triangular region in parameter space with an area dependent on the number of qubits used, shown in Fig.~\ref{fig:lyap_diagram}. From this it was predicted that a minimum of six qubits is necessary to observe Lyapunov decay in the quantum sawtooth map, independent of hardware platform. While other quantum maps were not analyzed closely, they are largely expected to require at least six qubits and even lower error per gate to reach the Lyapunov regime.

Lastly, numerical simulations and experimental data \cite{porter2021slowed} were employed to chart a path forward. While numerical results showed concretely how six or more qubits could exhibit Lyapunov decay, scaling arguments were employed to understand the relative importance of improving architecture versus reducing noise in Fig.~\ref{fig:abc_scaling}.

For IBM-Q it was found that $2-70\times$ less error per \code{CNOT} gate, including crosstalk, may enable an observation of Lyapunov decay on no fewer than six qubits. This range captures tradeoffs in future architectures, most prominently connectivity and gate parallelization, as well as uncertainty in the magnitude of crosstalk and the scalability of software-based optimization. For IonQ it was estimated that $0.4-80\times$ less error per M{\o}lmer-S{\o}rensen gate may enable an observation of Lyapunov decay. This range captures the possibility of parallel gates and uncertainty around the magnitude of dynamical and crosstalk effects.

For more than six qubits the larger Hilbert space lowers the $1/N$ fidelity limit which allows for faster initial decay, resulting in surprisingly accessible scaling. This is only possible due to the noise-resilient nature of observing the Lyapunov exponent. This demonstrates the efficiency of using fidelity to extract useful quantum information and achieve quantum advantage.

While Sec.~\ref{sec:requirements} only considered IBM-Q's superconducting architecture and IonQ's trapped ion architecture, there are many promising quantum testbeds coming online for which a similar analysis can be performed. Other near term platforms that use superconductors, trapped ions, neutral atoms, photons, or other qubit types may have more favorable trade-offs in terms of connectivity, parallelization, crosstalk, and other factors that could help to reach the Lyapunov regime more quickly. One example of such a trade-off is the optimal control approach used by the LLNL Quantum Design and Integration Testbed (QuDIT) \cite{wu2020high, holland2020optimal}, which computes waveforms for arbitrary gates to reduce the gate depth. This was shown to outperform Rigetti by an order of magnitude in simulation duration, in proportion to the length of Rigetti's native gate decomposition \cite{shi2021simulating}.

Out-of-time-ordered correlators (OTOCs) are a quantum analogue to the classical Lyapunov exponent that have been widely explored recently, but were not considered in this work. They allow probing of exponential information scrambling in quantum-chaotic systems \cite{rozenbaum2017lyapunov, larkin1969quasiclassical}. Recent work suggests deep connections between the fidelity (Loschmidt echo) and OTOCs \cite{yan2020information}. Both could potentially detect quantum chaos on near-term quantum devices, and OTOCs were even noted recently to show an exponential trend on fewer qubits than the fidelity \cite{pg2021out}. However the OTOC growth rate is known to differ in several ways from the Lyapunov exponent \cite{rozenbaum2017lyapunov} so the fidelity may be more reliable for quantum-classical correspondence. Additionally, the fidelity is perfectly suited to observing the Lyapunov exponent in the presence of large noise, whereas the study of quantum-classical correspondence of OTOCs in the presence of noise is still developing \cite{chavez2019quantum, goldfriend2020quasi}.

In addition to the single-body quantum chaos studied here, simulating many-body quantum chaos is also of great interest. A popular choice for this is quantum spin chains with local interactions, partly due to their natural mapping to qubits with local gates. However quantum maps can also be generalized to many-body systems by adding local coupling terms in the potential energy. One example is the classical many-body kicked rotor \cite{rajak2018stability}, which has a corresponding quantum Hamiltonian \cite{notarnicola2020slow} and therefore a quantum map. This allows simulation of many-body quantum chaos without need for trotterization, potentially realizing interesting dynamics in fewer time steps. The main questions are whether this gain is undone by the need for a quantum Fourier transform and whether efficient algorithms for such systems exist.

The fields of quantum and classical chaos are ripe for exploration via quantum simulation, and hopefully the objective outlined here will be the first of many similar advances.

\section{Acknowledgements}

The authors thank the Quantum Leap group at LLNL for stimulating discussions and ideas that improved the manuscript, including Vasily Geyko, Frank R. Graziani, Stephen B. Libby, Roger W. Minich, and Yuan Shi. We also thank Kristin M. Beck, Alessandro R. Castelli, Jonathan L. DuBois, and Yaniv J. Rosen of the Quantum Coherent Device Physics group at LLNL and Robert Tyler Sutherland at UT San Antonio for their insights into quantum hardware and noise processes. We especially thank Kristin M. Beck for lending her expertise on the IonQ platform. We thank Philippe Jacquod at the University of Geneva for clarifying discussion of the bandwidth limit.

This work was performed by LLNL under the auspices of the U. S. DOE under Contract DE-AC52-07NA27344 and was supported by the DOE Office of Fusion Energy Sciences “Quantum Leap for Fusion Energy Sciences” project FWP-SCW1680 and by LLNL Laboratory Directed Research and Development project 19-FS-078.

\appendix

\section{Effect of noise on the classical sawtooth map}

\begin{figure}
	\centering
	\includegraphics[width=80mm]{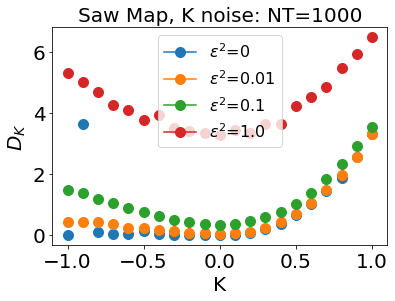}
	\caption{Classical diffusion dependence on the variance $\epsilon^2$ of random noise in parameter $\Kparam$.}
	\label{fig:sims_clas_noise}
\end{figure}

Noise in a classical sawtooth map can be understood through its effect on diffusion. For the random noise in parameter $\Kparam$ considered in this paper the noise itself can dominate the diffusion, potentially affecting the localization condition in the quantum system. As shown in Fig.~\ref{fig:sims_clas_noise} the effect of noise is greatest in the quasi-integrable regime $-4<\Kparam<0$, but still significant in the chaotic regime $\Kparam>0$. When the dynamics are chaotic these results can be understood through a random phase approximation
\begin{flalign}
	D_{\Kparam, \epsilon} &\approx \langle (\Delta J)^2 \rangle = \frac{1}{2\pi} \int_{-\pi}^{\pi} d\theta \langle (K+\xi)^2 \rangle \theta^2 \nonumber \\
	&= \frac{\pi^2}{3} (\Kparam^2 + \epsilon^2)
	\label{eq:appendix_diffusion}
\end{flalign}
Different dependence can occur for different types of noise. For instance noise directly in the update rule for $\theta$ has a negligible effect on diffusion in the chaotic regime, as shown for the standard map in Fig.~5.20 of Lichtenberg and Lieberman \cite{lichtenberg1992regular}.

Note that one can use $\epsilon=\sigma*2\pi\Lparam/N$ to convert $\sigma$ to $\epsilon$ in the Lyapunov regime, with $\Lparam=1$ assumed to maximize the extent of the Lyapunov regime. For the largest $\sigma$ that allows detection of the Lyapunov regime (see Figs.~\ref{fig:sims_fid_decay_and_slope_dif_n=10} and \ref{fig:lyap_decay_n=5,6}) at various $n \le 12$ one finds $\epsilon^2<0.01$, suggesting a negligible effect of noise on the diffusion for $\Kparam \gg 0.1$.

\bibliographystyle{unsrt}
\bibliography{qsm}

\end{document}